\documentclass[12pt]{article}

\font\grande=cmr9.5 scaled \magstep4
\font\medio=cmr9.5 scaled \magstep2
\outer\def\beginsection#1\par{\medbreak\bigskip
      \message{#1}\leftline{\bf#1}\nobreak\medskip
\vskip-\parskip
      \noindent}
\usepackage{graphicx} 
\begin{document}
\bibliographystyle {unsrt}

\titlepage

\vspace*{10mm}
\begin{center}
{\grande Quasiadiabatic modes from viscous inhomogeneities}\\
\vspace{15mm}
 Massimo Giovannini 
 \footnote{Electronic address: massimo.giovannini@cern.ch} \\
\vspace{0.5cm}
{{\sl Department of Physics, Theory Division, CERN, 1211 Geneva 23, Switzerland }}\\
\vspace{1cm}
{{\sl INFN, Section of Milan-Bicocca, 20126 Milan, Italy}}
\vspace*{1cm}

\end{center}

\vskip 1cm
\centerline{\medio  Abstract}
\vskip 1cm
The viscous inhomogeneities of a relativistic plasma determine a further class of
entropic modes whose amplitude must be sufficiently small since curvature perturbations are observed to be predominantly 
adiabatic and Gaussian over large scales. When the viscous coefficients only depend on the energy density of the fluid the 
corresponding curvature fluctuations are shown to be almost adiabatic.  
After addressing the problem in a gauge-invariant perturbative expansion, the same analysis 
is repeated at a non-perturbative level by investigating the nonlinear curvature inhomogeneities induced by the spatial variation of the viscous coefficients.  It is demonstrated that the quasiadiabatic modes are suppressed in comparison with a bona fide adiabatic solution.
Because of its anomalously large tensor to scalar ratio the quasiadiabatic mode cannot be a substitute for the
conventional adiabatic paradigm so that, ultimately, the present findings seems to exclude the possibility of a successful 
accelerated dynamics solely based on relativistic viscous fluids.  If the dominant adiabatic mode is not affected by the viscosity of the background
a sufficiently small fraction of entropic fluctuations of viscous origin cannot be a priori ruled out. 

\noindent

\vspace{5mm}

\vfill
\newpage
\renewcommand{\theequation}{1.\arabic{equation}}
\setcounter{equation}{0}
\section{Introduction}
\label{sec1}
The first releases of the WMAP data \cite{WMAP1} presented convincing evidence that the initial conditions of the Einstein-Boltzmann hierarchy are predominantly adiabatic. This conclusion follows from the relative position of the first anticorrelation peak of the cross-spectrum between the temperature and the polarization of the Cosmic Microwave Background (CMB in what follows). The subsequent WMAP releases and the Planck explorer results \cite{WMAP2} confirmed (and refined) the early determinations of the first WMAP data \cite{WMAP1} so that today we can say, with a fair degree of confidence, that  single field inflationary models lead naturally to adiabatic initial data of the CMB temperature and polarization anisotropies. Nonetheless every deviation from this simple paradigm leads necessarily to various entropic solutions (see e.g.\cite{hh1,hh2}).  The entropic modes can be explicitly constrained using CMB physics but their presence is not mandatory 
for a consistent explanation of the observational data. Conversely the presence of an adiabatic mode is essential 
and cannot be overlooked at least in the framework of the standard concordance paradigm. According to the current data \cite{WMAP1,WMAP2}, a small fraction of anticorrelated entropic modes in the presence of a dominant adiabatic mode may even improve the fit of the temperature autocorrelations accounting for potential large-scale suppressions of the corresponding angular power spectra.

The conventional distinction between the adiabatic and the entropic solutions, going back to the pioneering analyses of the temperature and polarization anisotropies \cite{pee1,pee2},  assumes an ambient fluid that is thermodynamically reversible but this hypothesis is not necessary and it can be relaxed by making the plasma viscous. The gauge-invariant perturbations of the viscous coefficients lead to new fluctuations modes of the predecoupling plasma \cite{mg1}. The physical features of these viscous solutions differ from the four conventional nonadiabatic modes\footnote{The four nonadiabatic modes are customarily classified into baryon-radiation, CDM-radiation, neutrino velocity and neutrino entropy modes \cite{hh1,hh2}.}. Since the inhomogeneities of the viscous coefficients cause entropic fluctuations of the spatial curvature, the r\^ole of viscosity at large scales must either be constrained by the initial data 
of the Einstein-Boltzmann hierarchy \cite{mg1} or totally absent. If correct this conclusion would threaten the possibility of an accelerated phase only driven by the viscous coefficients. In this paper we shall therefore analyze the fluctuations induced by the viscous coefficients both at the linear and at the nonlinear level. It will be shown that the large-scale fluctuations induced by inhomogeneous viscosities are not necessarily entropic, as argued in the previous paragraph, but they can be very close to adiabatic at large scales (hence the terminology quasiadiabatic) provided the viscous coefficients solely depend on the energy density of the relativistic plasma. The  evolution equations of the gauge-invariant curvature perturbations in the case of a relativistic, irrotational and irreversible fluid differ substantially from the ones valid in the perfect fluid case and derived by Lukash  \cite{lukash} even prior to the formulation of the Bardeen formalism \cite{bard}. The equations for the normal modes of the curvature perturbations driven by the viscous inhomogeneities reduce anyway to the corresponding expression valid for a perfect and relativistic fluid \cite{lukash}  when the viscous coefficients vanish exactly.  The results obtained in perturbation theory (within a gauge-invariant approach) are confirmed by a  fully nonlinear analysis where the curvature inhomogeneities are treated within the expansion in spatial gradients. In this approach the curvature inhomogeneities are not required to be perturbative. 

If the inflationary phase is solely driven by the viscous coefficients the quasiadiabatic scalar mode is more suppressed 
 than in the conventional case where inflation is driven by a single scalar field.  Consequently the tensor to scalar ratio exceeds the observational limits and turns out to be excessively large if compared with a bona fide adiabatic solution.
 Viscous stresses have been widely studied already in the mid seventies since 
 the presence of bulk viscosity in the relativistic plasma can influence the character of the cosmological 
singularity \cite{bel1}. Barrow \cite{bv1} presented detailed studies 
 on inflationary Universes driven by a bulk viscosity coefficient (see also \cite{bv1a,bv2}). 
While the  possibility of an inflationary phase only driven by bulk viscosity received 
various critiques also in the past (see e.g. second paper of Ref. \cite{bv1}) one of the byproducts 
of the present analysis is a systematic strategy for a more concrete phenomenological 
scrutiny of the large scale inhomogeneities induced by the viscous coefficients. 
 
The plan of this paper is therefore the following. In section \ref{sec2} we shall describe the viscous fluctuations of  a relativistic plasma in an 
explicitly gauge-invariant language. Section \ref{sec3} is devoted to the normal modes of the system; we shall also investigate if and when the perturbative fluctuations of the bulk viscosity can become quasiadiabatic. In section \ref{sec4} the themes scrutinized in section \ref{sec3} will 
be examined in a fully nonlinear perspective by expanding the geometry and the hydrodynamical variables in  spatial gradients.
In section \ref{sec5} the tensor to scalar ratio will be computed when the quasi-de Sitter phase is solely driven by bulk viscosity. Finally section \ref{sec6} contains some concluding remarks. 

\renewcommand{\theequation}{2.\arabic{equation}}
\setcounter{equation}{0}
\section{Gauge-invariant fluctuations viscous coefficients}
\label{sec2}
\subsection{General considerations}
In what follows we shall present the full governing equations for an irreversible fluid 
where the viscous coefficients can fluctuate in space and time.
If the total energy-momentum tensor of the plasma includes the viscous contributions, the adiabatic limit, in a strict sense, is recovered when the viscosities are neglected and the total entropy four-vector is conserved.  The total energy-momentum tensor of the fluid shall then be written, for the present ends, as:
\begin{eqnarray}
{\mathcal T}_{\mu}^{\nu} &=& (p + \rho) \,\,u_{\mu} \,u^{\nu} - p\delta_{\mu}^{\nu} + \xi\,\,{\mathcal P}_{\mu}^{\nu} \,\,\nabla_{\alpha} u^{\alpha} + 2 \,\,\eta\,\, \sigma_{\mu}^{\nu},
\nonumber\\
\sigma_{\mu}^{\nu} &=& \frac{1}{2} {\mathcal P}_{\mu}^{\alpha} \,\,{\mathcal P}_{\beta}^{\nu}\,\, \biggl[ \nabla_{\alpha} u^{\beta} + \nabla^{\beta} u_{\alpha} -\frac{2}{3} \delta^{\alpha}_{\beta} (\nabla_{\lambda} u^{\lambda}) \biggr].
\label{BVC1}
\end{eqnarray}
where, as usual,  ${\mathcal P}_{\mu}^{\nu}= \biggl( \delta_{\mu}^{\nu} - u_{\mu} u^{\nu} \biggr)$, $\xi$ denotes the 
bulk viscosity coefficient, $\eta$ the shear viscosity and four-velocity obeys  $g^{\mu\nu} \, u_{\mu} \, u_{\nu} =1$. 
We shall consider the situation where the bulk viscosity coefficient $\xi$ is the sum of a homogeneous part (denoted by $\overline{\xi}(\tau)$) supplemented by the inhomogeneous contribution (denoted by $\delta\xi(\tau, \vec{x})$):
\begin{equation}
\xi(\tau,\vec{x}) = \overline{\xi}(\tau) + \delta \xi(\tau,\vec{x}).
\label{decomp}
\end{equation}
Note that the fluctuations of the bulk viscosity are defined in a slightly different manner\footnote{In the first paper of Ref. \cite{mg1} the bulk viscosity 
 coefficient includes a supplementary scale factor. More precisely we have that  $a\,\xi_{old} = \xi_{current}$ 
 where $\xi_{old}$ and $\xi_{current}$ 
 correspond, respectively, to the previous and to the current definitions of the bulk viscosity coefficients.}  in comparison with Ref. \cite{mg1}. 
The fluctuations of the bulk viscosity coefficient would read $\eta(\tau, \vec{x}) = \overline{\eta}(\tau) + \delta\eta(\tau, \vec{x})$.
In what follows we shall only consider the fluctuations of the bulk viscosity coefficient. A potential 
 homogeneous component of the shear viscosity coefficient would not contribute the background equations (\ref{FL}); 
 indeed $\sigma_{\mu}^{\nu}$ of Eq. (\ref{BVC1}) does not have a homogeneous component, as it can be explicitly verified.  
The shear viscosity does not contribute to the background so that, without loss of generality, we can take $\overline{\eta} \to 0$.
 The shear viscosity is therefore important only over small scales. Indeed across the matter-radiation transition the shear viscosity coefficient $\eta$  determines the optical depth, the Silk damping scale and, ultimately, the shape of the visibility function \cite{pee1,visibility}. Over large scales (possibly exceeding the Hubble radius) the shear viscosity only couples to the traceless part of the extrinsic 
curvature\footnote{To be precise we refer here to the extrinsic curvature of the spatial slices called $K_{ij}(\vec{x},\tau)$; 
see, in this respect, the discussion of section \ref{sec4}.}.

The viscous energy-momentum tensor at large-scales can be evaluated in the Landau-Lifshitz frame \cite{vis1}. The covariant conservation of the total energy-momentum tensor (i.e. $\nabla_{\mu} {\mathcal T}^{\mu\nu}=0$) can be 
projected along $u_{\nu}$ and along ${\mathcal P}_{\nu}^{\alpha}$; the two obtained equations together with the covariant conservation of the particle current are given hereunder:
\begin{eqnarray}
&& \nabla_{\mu}[ (p_{t} + \rho_{t}) u^{\mu}] - u_{\alpha} \partial^{\alpha} p_{t} + u_{\beta} \nabla_{\alpha} {\mathcal T}^{\alpha\beta} =0,
\label{c1}\\
&& (p_{t} + \rho_{t}) u^{\beta} \nabla_{\beta} u^{\alpha} - \partial^{\alpha}p_{t} + u^{\alpha} u_{\beta} \partial^{\beta} p_{t} + {\mathcal P}^{\alpha}_{\nu} \nabla_{\mu} {\mathcal T}^{\mu\nu} =0,
\label{c2}\\
&& \nabla_{\alpha}( n_{t} u^{\alpha} + \nu^{\alpha}) =0.
\label{c3}
\end{eqnarray}
Using then Eqs. (\ref{c1}) and (\ref{c3}) together  
with the first principle of thermodynamics, the evolution of the entropy\footnote{The explicit form of Eq. (\ref{entro}) has been obtained by trading the term $u_{\nu} \nabla_{\mu} {\mathcal T}^{\mu\nu}$ for 
$(\nabla_{\mu} u_{\nu}) {\mathcal T}^{\mu\nu}$ since, in the Landau frame, $\nabla_{\mu} ( u_{\nu} {\mathcal T}^{\mu\nu} ) = 0$.} can be easily derived:
\begin{equation}
\nabla_{\alpha} [ s u^{\alpha} - \overline{\mu} \nu^{\alpha} ] + \nu^{\alpha} \partial_{\alpha} \overline{\mu} =
\xi (\nabla_{\alpha} u^{\alpha})^2/T + 2 \eta \, \sigma_{\mu\nu} \,\sigma^{\mu\nu}/T,
\label{entro}
\end{equation}
where the right hand side of Eq. (\ref{entro}) comes directly from $ \nabla_{\alpha} u_{\beta} \, {\mathcal T}^{\alpha\beta}/T$;
in Eq. (\ref{entro}) $\overline{\mu} = \mu/T$ is the chemical potential rescaled through the temperature, $s$ is the entropy density and $\nu_{\alpha}$ is given by:
\begin{equation}
\nu_{\alpha} = \chi \biggl(\frac{n_{t} T}{\rho_{t} + p_{t}}\biggr)^2 \biggl[ \partial_{\alpha} \overline{\mu} - u_{\alpha} u^{\beta}\partial_{\beta} \overline{\mu} \biggr],
\label{nunu}
\end{equation}
where $\chi$ denotes the heat transfer coefficient. The adiabatic limit is recovered when the viscous contributions are neglected and the total entropy four-vector is conserved.

\subsection{Metric fluctuations induced by bulk viscosity} 

We shall now derive the gauge-invariant system of metric fluctuations with the purpose of computing the curvature 
perturbations induced by the fluctuations of the viscous coefficients. In section \ref{sec4} this problem will be addressed in fully nonlinear terms, i.e. 
without relying on the separation of the various quantities into a background value supplemented by the corresponding fluctuations. 
For the time being the metric, the total energy density and the total pressure will be split as
\begin{eqnarray}
\rho(\tau, \vec{x}) &=& \rho_{t}(\tau) + \delta_{s} \rho_{t}(\tau, \vec{x}), \qquad p(\tau, \vec{x}) = p_{t}(\tau) + \delta p_{t}(\tau, \vec{x}), 
\nonumber\\
g_{\mu\nu}(\tau, \vec{x}) &=& \overline{g}_{\mu\nu}(\tau) + \delta g_{\mu\nu}(\tau, \vec{x}),
\end{eqnarray}
where $\overline{g}_{\mu\nu}$ will be taken to be conformally flat, i.e. $\overline{g}_{\mu\nu} = a^2(\tau) \eta_{\mu\nu}$ where $a(\tau)$ is the scale factor, $\tau$ is the conformal time coordinate and $\eta_{\mu\nu}$ is the Minkowski metric with signature mostly minus. In general 
$\delta g_{\mu\nu}$ to linear order can always be written as the sum the scalar, tensor and vector fluctuations, i.e. 
$\delta g_{\mu\nu} = \delta_{s} g_{\mu\nu} + \delta_{t} g_{\mu\nu} + \delta_{v} g_{\mu\nu}$, where $\delta_{s}$, $\delta_{t}$ and 
$\delta_{v}$ denote, respectively, the scalar tensor and vector fluctuations of the perturbed metric.

Using Einstein equations and recalling the notations of Eqs. (\ref{BVC1})--(\ref{decomp}) the evolution for the homogeneous expansion rate are\footnote{The Planck length and the Planck mass will be defined, respectively, as $\ell_{P} =/1\sqrt{8 \pi G} = 1/\overline{M}_{P}$ where $\overline{M}_{P} = M_{P}/\sqrt{8 \pi}$. Recall also that $M_{P} = 1.22\times 10^{19}\,\, \mathrm{GeV}$}:
\begin{equation}
2 ({\mathcal H}^2 - {\mathcal H}') = \ell_{P}^2 a^2 (\rho_{\mathrm{t}} + {\mathcal P}_{\mathrm{t}}), \qquad 3 {\mathcal H}^2 = 
\ell_{P}^2 a^2 \rho_{\mathrm{t}},\qquad {\mathcal H} = \frac{a^{\prime}}{a},
 \label{FL}
 \end{equation}
 where ${\mathcal P}_{t}= p_{t} - 3 \overline{\xi} \, {\mathcal H}/a$  is the shifted background pressure. In Eq. (\ref{FL}) 
 the prime denotes a derivation with respect to $\tau$ while the overdot will denote a derivation with respect 
 to the cosmic time coordinate $t$. The relation between ${\mathcal H}$ and $H=\dot{a}/a$ is given, as usual, by 
 ${\mathcal H} = a H$.   It is appropriate to remark, at this point, that the perfect fluid contribution is characterized, in the simplest case, by 
the barotropic index $w= p_{t}/\rho_{t}$ and by the related sound speed $c_{s}^2 = p_{t}^{\prime}/\rho^{\prime}$. 

The scalar fluctuations of the conformally flat metric 
are parametrized by four independent functions so that the entries of the perturbed metric can be written as: 
\begin{equation}
\delta_{s} g_{00} = 2 a^2 \phi,\qquad  \delta_{s} g_{ij} = 
2 a^2 (\psi \delta_{i j} - \partial_{i} \partial_{j} \alpha),\qquad \delta_{s} g_{0i} = - a^2  \partial_{i} \beta,
\label{scalar}
\end{equation}
where, as already mentioned, $\delta_{s}$ denotes the scalar fluctuation of the perturbed metric.
The tensor modes are immediately gauge-invariant and are parametrized in terms 
of a divergenceless and traceless rank-two tensor in three dimensions:
\begin{equation}
\delta_{t} g_{i j} = - a^2 h_{ij},\qquad \partial_{i} h^{i}_{j} =0, \qquad h_{i}^{i} =0,
\label{tensor}
\end{equation}
where, as already mentioned, $\delta_{t}$ denotes the tensor fluctuation of the perturbed metric.
Finally the vector modes are parametrized as
\begin{equation}
\delta_{v} g_{0 i} = - a^2 Q_{i},\qquad  \delta_{v} g_{i j} =  a^2 ( \partial_{i} W_{j} + \partial_{j} W_{i}), \qquad \partial_{i} Q^{i} = \partial_{i} W^{i} =0,
\label{vector}
\end{equation}
where, as already mentioned, $\delta_{v}$ denotes the vector fluctuation of the various entries of the perturbed metric.

\subsection{Evolution of the tensor and vector modes}
The evolution of the tensor and vector modes is completely standard. In particular the tensor modes obey, as usual, 
\begin{equation}
h_{ij}^{\prime\prime} + 2 {\mathcal H} h_{ij}^{\prime} - \nabla^2 h_{ij} =0.
\label{TT}
\end{equation}
Equation (\ref{TT}), as expected, does not include a direct contribution of the bulk viscosity coefficient. The contribution is only 
indirect (i.e. through the scale factor). From Eqs. (\ref{vector}) and recalling that $\overline{u}_{0}\delta_{ v} u^{i} = V^{i}$,  
the $(0i)$ and $(ij)$ components of the perturbed Einstein equations together with the evolution 
of the velocity give
\begin{eqnarray}
&& \nabla^2 Q_{i} = - 2 \ell_{P}^2 ( \rho_{t} + {\mathcal P}_{t}) a^2  V_{i},\qquad Q_{i}' + 2 {\cal H} Q_{i} =0, 
\label{VP5}\\
&& \bigl[   V_{ i} (\rho_{t} + {\mathcal P}_{t})\bigr]^{\prime} + 4 {\cal H} \bigl[  V_{i} ( \rho_{t} + {\mathcal P}_{t} ) \bigr]=0,
\label{VP6}
\end{eqnarray}
where, for simplicity, we just selected the gauge $W_{i}=0$.
Equation (\ref{TT}) and its solution will play a r\^ole in the determination of the tensor to scalar ratio to be specifically discussed 
later on. Equations (\ref{VP5}) and (\ref{VP6}) determine the rate of decrease of the vector modes which do not play a major r\^ole 
in the present investigation,  exactly as in the other more conventional cases. Equations (\ref{TT}), (\ref{VP5}) and (\ref{VP6}) 
share a common feature: the effect of the viscosity enters only through the evolution of the background. This is in sharp 
contrast with what happens in the case of the scalar modes.

\subsection{Evolution of the scalar modes}
The scalar fluctuations of the effective energy-momentum tensor\footnote{From now on we shall omit 
the subscript and write $\delta \rho_{t}$ (instead of $\delta_{s} \rho_{t}$), $\delta p$ (instead of $\delta_{s} p_{t}$) 
and so on and so forth.} are given by:
\begin{eqnarray}
 \delta_{s} {\mathcal T}_{0}^{0} &=& \delta \rho,\qquad \delta_{s} {\mathcal T}_{0}^{i} 
 =  \biggl( \rho_{t} + p_{t} - 3 \frac{\overline{\xi}}{a} {\cal H} \biggr) v^{i},
 \label{T00T0i}\\
 \delta_{s} {\mathcal T}_{i}^{j} &=& - \delta_{i}^{j}\biggl\{ \delta p_{t} - 
 3 \frac{{\cal H}}{a} \delta \xi - \frac{\overline{\xi}}{a} \biggl[ \theta - 3 
 ( \psi' + {\cal H} \phi) + \nabla^2 \alpha'\biggr]\biggr\},
 \label{PT2}
 \end{eqnarray}
 where $u_{0} \delta_{s} u^{i} = v^{i}$, and $\partial_{i} v^{i} = \theta$. 
 Concerning Eqs. (\ref{T00T0i}) and (\ref{PT2}) two comments are in order: first, as expected, the bulk viscosity and its fluctuations affect the spatial components of the perturbed energy-momentum tensor; second the bulk viscosity couples both to the peculiar velocity and to 
the metric fluctuations. Consequently the inhomogeneities of the bulk viscosity cannot be rescaled away 
with simple redefinitions of the metric perturbations as in the case of Eqs. (\ref{TT}), (\ref{VP5}) and (\ref{VP6}).
 
Equations (\ref{T00T0i})--(\ref{PT2}) are written in general terms since no particular gauge choice has been imposed so far. 
The same strategy will be also used in all the other relevant equations with the aim of deriving a 
consistent gauge-invariant evolution involving also the viscous coefficients and their fluctuations.  
Since this procedure is a bit lengthy but only the main steps will be swiftly outlined by focussing on the gauge-invariant 
meaning of the bulk viscosity coefficients. Without imposing a specific gauge choice, the Hamiltonian and the momentum constraints
stemming from the $(00)$ and $(0i)$ perturbed Einstein equations in the presence of the viscous energy-momentum tensor 
are given by: 
\begin{eqnarray}
&& \nabla^2 \psi - 
{\mathcal H} \nabla^2( \beta - \alpha') - 3 {\cal H} ( \psi' + {\mathcal H} \phi) = \frac{\ell_{P}^2}{2} a^2 \delta \rho_{t},
\label{PE00}\\
&& \nabla^2 ( \psi' + {\mathcal H} \phi) + ( {\mathcal H}^2 - {\cal H}') (\nabla^2 \beta  + \theta) =0.
\label{PE0i}
\end{eqnarray}
The trace of the perturbed spatial components of the Einstein equations  is:
\begin{eqnarray}
&&  \psi''  +({\cal H}^2 + 2 {\cal H}') \phi + {\cal H} (\phi' + 2 \psi') + 
\frac{1}{2}
\nabla^2 [ ( \phi - \psi) + ( \beta - \alpha')' + 2 {\cal H} ( \beta - \alpha') ]
\nonumber\\
&&= \frac{\ell_{P}^2}{2} a^2\biggl\{ \delta p_{t} - 3 \frac{ {\cal H}}{a} \delta \xi - \frac{\overline{\xi}}{a}  
\biggl[ \theta - 3 (\psi' + {\cal H} \phi) + \nabla^2 \alpha'\biggr]\biggr\}.
\label{Pij}
\end{eqnarray}
Similarly the traceless projection of the spatial components becomes:
\begin{equation}
{\mathcal L}_{i}^{j} \biggl[( \phi - \psi) -  ( \alpha' - \beta)' + 2 {\mathcal H} ( \alpha' - \beta) \biggr]  = \ell_{P}^2\, a^2\,  \Pi_{i}^{j},
\label{Pineqj}
\end{equation}
where ${\mathcal L}_{i j} =(\partial_{i}\partial_{j}  - \delta_{i j} \nabla^2/3)$  and $\Pi_{ij}$ is the total 
anisotropic stress. Note that $\Pi_{ij}$ is gauge-invariant and this implies that also 
the left hand side must be gauge-invariant\footnote{ Note that in the subsequent applications we shall
use the following notation $\partial_{i} \partial_{j} \Pi^{ij} = \nabla^2 \Pi$. }. This observation essentially determines the form of the so-called 
Bardeen potentials \cite{bard} $\Phi = \phi + ( \beta - \alpha')' + {\mathcal H} ( \beta - \alpha')$ 
and $\Psi= \psi - {\mathcal H} ( \beta - \alpha')$; consequently  the gauge-invariant counterpart of  Eq. (\ref{Pij}) 
is\footnote{ This shows, once more, that the perturbed equations cannot be simply obtained by rescaling from the conventional situation.}
\begin{equation}
\Psi'' + {\cal H} ( \Phi' + 2 \Psi') + ( {\cal H}^2 + 2 {\cal H}') \Phi = \frac{\ell_{P}^2}{2}a^2 
\biggl\{ \delta p_{g} - \frac{3 {\cal H}}{a} \Xi - \frac{\overline{\xi}}{a}\biggl[ \Theta - 3 ( \Psi' + 
{\cal H} \Phi)\biggr]\biggr\},
\label{Gij}
\end{equation}
where  $\Xi$ denotes the gauge-invariant fluctuation of the bulk viscosity coefficient  defined as: 
\begin{equation}
\Xi = \delta \xi + \overline{\xi}' ( \beta - \alpha').
\label{XI}
\end{equation}
The gauge-invariant expression of the viscosity fluctuation will be particularly important in section \ref{sec3} when discussing 
the quasiadiabatic mode. 
In Eq. (\ref{Gij}) we also introduced the gauge-invariant counterparts of $\delta p_{t}$ and $\theta$ namely
\begin{eqnarray}
\delta \rho_{g} = \delta \rho_{t} + \rho_{t}' ( \beta - \alpha'), \qquad \delta p_{g} = \delta p_{t} + p_{t}' ( \beta - \alpha'),
\qquad \Theta = \theta + \nabla^2 \alpha',
\label{fluid}
\end{eqnarray}
where the subscript $g$ in Eq. (\ref{fluid}) recalls that the corresponding fluctuation is  is invariant under infinitesimal coordinate transformations (for short gauge-invariant). 
The gauge-invariant counterpart of Eqs. (\ref{PE00}) and (\ref{PE0i}) becomes respectively
\begin{eqnarray}
\nabla^2 \Psi - 3 {\cal H} ( \Psi' + {\cal H} \Phi) = 
\ell_{P}^2 a^2 \delta\rho_{g}/2,\qquad  \nabla^2 ( \Psi' + {\cal H} \Phi) + ({\cal H}^2 - {\cal H}')\Theta = 0 , 
\label{G00G0i}
\end{eqnarray}
Equation (\ref{Pineqj}) implies instead 
\begin{equation}
\nabla^4 (\Phi - \Psi) = 3 \ell_{P}^2\, a^2\, \nabla^2 \Pi/2, \qquad \partial_{i} \partial_{j} \Pi^{ij} = \nabla^2 \Pi.
\label{diff}
\end{equation}
The evolution of the total energy-momentum tensor implies the following pair of equations
\begin{eqnarray}
&& \delta \rho_{g}'+ 3 {\cal H} ( \delta \rho_{g} + \delta p_{g} )
 + ( \rho_{t} + {\mathcal P}_{t}) \Theta - 3 ( \rho_{t}+ {\mathcal P}_{t}) \Psi'  =  {\mathcal F}_{\xi},
\label{DRGITOT}\\
&&  \Theta' + \biggl[  
4 {\cal H} + \frac{( \rho_{t}' + {\mathcal P}_{t}')}{ \rho_{t} + {\mathcal P}_{t}}\biggr]\Theta+ 
\frac{\nabla^2 \delta p_{g}}{\rho_{t} + {\mathcal P}_{t}}+ \nabla^2 \Phi 
= {\mathcal G}_{\xi}, 
\label{THGITOT}
\end{eqnarray}
where the two source terms ${\mathcal F}_{\xi}$ and ${\mathcal G}_{\xi}$ are defined, respectively, as:
\begin{eqnarray}
 {\mathcal F}_{\xi} &=& 9 \,\frac{{\cal H}}{a}\, \Xi +  \frac{3}{a} {\mathcal H} \overline{\xi} [ 
\Theta - 3 ( \Psi' + {\mathcal H} \Phi)],
\label{Fxi}\\
 {\mathcal G}_{\xi} &=& \frac{3\,\,{\mathcal H} \,\,\nabla^2 \Xi}{a (\rho_{t} + {\mathcal P}_{t}) }
+ \frac{\overline{\xi}\,\, [ \nabla^2 \Theta - 3 \nabla^2(\Psi' + {\mathcal H} \Phi)]}{a (\rho_{t} + {\mathcal P}_{t})}.
\label{Gxi}
\end{eqnarray} 
Both ${\mathcal F}_{\xi}$ and ${\mathcal G}_{\xi}$ contain the bulk viscosity and its fluctuations.
As we shall see in the following section the presence of spatial fluctuations 
in the bulk viscosity coefficient induces further source terms in the evolution of the spatial curvature. As we shall see these terms 
play effectively the same r\^ole of an  intrinsic nonadiabatic pressure fluctuation. 

\renewcommand{\theequation}{3.\arabic{equation}}
\setcounter{equation}{0}
\section{Quasiadiabatic modes}
\label{sec3}
From the governing equations of the previous section the gauge-invariant evolution of the curvature perturbations
can be easily obtained. We shall show, as anticipated, that the spatial perturbations of the viscous coefficients act as a source term of the evolution equations of the curvature perturbations. They are physically equivalent to nonadiabatic pressure fluctuations. 
However, if the viscous coefficients depend solely on the energy density of the fluid, the source terms induced by the viscous coefficients 
can be neglected for typical scales larger than the Hubble radius. Over large-scales the evolution equations of the normal modes of the system reproduce the ones of the adiabatic modes. Conversely,  inside the Hubble radius their evolution equations are very different from their adiabatic counterpart. This is the reason why these modes have been termed here quasiadiabatic. 

\subsection{Adiabatic and nonadiabatic fluctuations of the pressure}
In terms of the Bardeen potentials the gauge-invariant curvature fluctuations are
\begin{equation}
{\mathcal R} = - \biggl[\Psi + \frac{{\mathcal H}}{{\mathcal H}^2 - {\mathcal H}'}\biggl({\mathcal H} \Phi + \Psi'\biggr) \biggr] = - ( \Psi - {\cal H} V_{\rm g}),
\label{defRL}
\end{equation}
where the second equality follows from the momentum constraint in its gauge-invariant form using the 
notation $\Theta = \nabla^2 V_{\rm g}$. In the comoving orthogonal gauge (where both $\beta$ and the three-velocity 
vanish) ${\mathcal R}$ coincides up to a sign (which is matter of conventions) with the fluctuations of the spatial curvature. 
While in  the comoving orthogonal gauge, ${\mathcal R}$ is related to fluctuations 
of the spatial curvature, in a different coordinate system ${\mathcal R}$ will have 
the same numerical value but will not necessarily 
be related to curvature fluctuations. Equation (\ref{defRL}) can be complemented with the definition of the curvature fluctuation on uniform 
density hypersurfaces:
\begin{equation}
\zeta = - \biggl( \Psi + {\mathcal H} \frac{\delta \rho_{g}}{\rho_{t}'}\biggr).
\label{defzetaL}
\end{equation}
Using Eq. (\ref{G00G0i}) the difference of Eqs. (\ref{defRL}) and (\ref{defzetaL}) 
is proportional to the Laplacian of the Bardeen potential and can therefore be neglected at large scales: 
 ${\mathcal R} = \zeta - 2\nabla^2 \Psi/[3 \ell_{P}^2 a ^2 ( \rho_{t} + {\mathcal P}_{t})]$
This equation generalizes the relation 
between ${\mathcal R}$ and $\zeta$ to the case when the bulk viscosity 
coefficient is non-vanishing and it implies that, up to Laplacians of $\Psi$, 
${\mathcal R} \simeq \zeta$.

Even if the conventional terminology might suggest otherwise, the nonadiabatic modes arise in a globally inviscid fluid, as the preceding considerations illustrate. In the present paper we want to drop this hypothesis since the total energy-momentum tensor of the plasma could include the contributions of the shear viscosity, of the bulk viscosity and of the heat transfer. The adiabatic limit, in a strict sense, is recovered when the viscous contributions are neglected and the total entropy four-vector is conserved.  Equation (\ref{defzetaL}) accounts for the 
curvature fluctuations on uniform density hypersurfaces. If the total fluid contains a number of different constituent 
components (for instance two, the $a$-fluid and the $b$-fluid) we can decompose the total $\zeta$ as $\zeta = \zeta_{a} (\rho_{a}'/\rho_{t}') + 
 \zeta_{b} (\rho_{b}'/\rho_{t}')$ where\footnote{Note that $\delta \rho_{g\,a}$ and $\delta\rho_{g\,b}$ are the gauge-invariant density 
fluctuations of the individual fluids and obviously $\rho_{t} = \rho_{a} + \rho_{b}$.}
\begin{equation}
\zeta_{a} = -\biggl( \Psi + {\cal H}\frac{\delta\rho_{g\,a}}{\rho_{a}'}\biggr),\qquad  \zeta_{b} =
 -\biggl( \Psi + {\cal H}\frac{\delta\rho_{g\,b}}{\rho_{b}'}\biggr).
\label{defzabL}
\end{equation}
While the weighted sum of $\zeta_{a}$ and $\zeta_{b}$ is related to the total $\zeta$, the difference between them gives 
what we normally define as the the entropy perturbations \cite{hh1,hh2}. More specifically, the relative 
fluctuations in the specific entropy $\varsigma$ can be written as\footnote{Note that $w_{a}$ and $w_{b}$ 
are the barotropic indices for the two fluids of the mixture.}:
\begin{equation}
{\mathcal S} = \frac{\delta \varsigma}{\varsigma} = -3 (\zeta_{\rm a} - \zeta_{\rm b}) \to \frac{\delta_{g\, a}}{( 1 + w_{a})} -  
\frac{\delta_{g\, b}}{( 1 + w_{b})}.
\label{entropydef}
\end{equation}
The the second equality in Eq. (\ref{entropydef}) holds in the case of two barotropic constituents. 
Equation (\ref{entropydef}) applies, for instance, in the discussion of CDM-radiation isocurvature
mode.  Let us now define the quantities relevant to the evolution 
of curvature perturbations. The gauge-invariant pressure fluctuation $\delta p_{g}$
can always be split into the adiabatic contribution (containing the total sound speed of the system) 
supplemented by the nonadiabatic contribution (containing the entropy fluctuations)
\begin{equation}
\delta p_{g} = \biggl(\frac{\delta p_{g}}{\delta \rho_{ g}}\biggr)_{\varsigma} \delta\rho_{ g}
+ \delta p_{nad}, \qquad \delta p_{nad}=
\biggl(\frac{\delta p_{ g}}{\delta \varsigma}\biggr)_{\rho} \delta\varsigma,
\label{deltapg}
\end{equation}
where the two subscripts imply that the two relative variations 
at the right-hand side should be taken, respectively, at constant entropy and energy densities\footnote{To perform the variation 
at constant (total) energy density means that
$\delta \rho_{g\, a} = - \delta\rho_{ g \,b }$. 
Similarly, to perform the variation at constant $\varsigma$ means that 
$\delta\varsigma =0$, i.e. from Eq. (\ref{entropydef}) $\delta\rho_{g\,a}/\rho_{\rm a}' =\delta \rho_{\rm g\,b}/\rho_{\rm b}'$.}.
In the simple case of two fluids the total speed of sound and the nonadiabatic
pressure density variation are:
\begin{equation}
c_{s}^2 =
 \biggl( \frac{\delta p_{g}}{\delta \rho_{g}}\biggr)_{\varsigma} = 
 \frac{c^2_{s\, a} \rho_{a}' + c^2_{s\,b} \rho_{ b}'}{\rho_{a}' + \rho_{ b}'},\qquad
  \delta p_{nad} =  \biggl( \frac{\delta p_{g}}{\delta \varsigma}\biggr)_{\rho} \delta \varsigma = 
- \frac{(c^2_{s\,a} - c^2_{s\,b}) \rho_{ a}' \rho_{b}'}{{\cal H} ( \rho_{a}' + 
\rho_{ b}')} (\zeta_{a} - \zeta_{b}),
\label{deltapnad1}
\end{equation}
where the speeds of sound in the two fluids of the mixture have been explicitly 
introduced. Recalling the connection between $\zeta$ and the weighted sum of
$\zeta_{\rm a}$ and $\zeta_{\rm b}$ it is also possible to write 
$\delta p_{nad} =  (c^2_{\rm s\,b} - c^2_{\rm s\,a}) \rho_{\rm a}' (\zeta_{\rm a} - \zeta)/{\mathcal H}$,
where the speeds of sound refer to the inviscid contribution to the total energy-momentum tensor. Thus 
$c^2_{s\, b} = w_{ b}$ and $c^2_{ s\, a} = w_{a}$.

\subsection{Decoupled evolution of the curvature perturbations}
The decoupled evolution of the curvature perturbations is obtained in two steps. 
The first step is to derive the first-order equation obeyed by the gauge-invariant curvature perturbations. The second step 
involves some lengthy but straightforward algebra to pass from the first-order (but still coupled) system to a second-order
decoupled equation. The first order equation obeyed by ${\mathcal R}^{\prime}$ can be obtained at least in two different ways 
either starting from the evolution of the metric perturbations or from the total velocity field. The first derivation 
consists in taking the difference of Eq. (\ref{Gij}) and of the Hamiltonian constraint (first equation of (\ref{G00G0i})). 
This combination will lead directly to a term containing 
a term proportional to $\delta p_{nad}$. The same result obtained with this procedure can be 
derived from Eq. (\ref{THGITOT}). In this case the observation is that, thanks to the momentum constraint (second 
equation of Eq. (\ref{G00G0i})), $\nabla^2({\mathcal R} + \Psi) = {\mathcal H} \Theta$: using this relation 
to eliminate $\Theta$ from Eq. (\ref{THGITOT}), the wanted equation can be immediately derived.  In both cases the final 
result can be expressed in the following manner:
\begin{equation}
{\mathcal R}' = \Sigma_{{\mathcal R}} - \frac{2{\mathcal H} c_{s}^2}{\ell_{P}^2 \,a^2(\rho_{t} + {\mathcal P}_{t})}\nabla^2 \Psi,
\label{evolR}
\end{equation}
where the total source term $\Sigma_{{\mathcal R}}$ is defined as
\begin{eqnarray}
\Sigma_{{\mathcal R}} &=& 
- \frac{{\cal H}}{\rho_{t} + {\mathcal P}_{t}} \delta p_{\rm nad}  + \frac{{\mathcal H}}{(\rho_{t} + {\mathcal P}_{t})} \Pi
\nonumber\\
&+& \frac{3 {\cal H}^2 }{a ( \rho_{t} + {\mathcal P}_{t})} \Xi   
+ \frac{3 {\mathcal H}}{a ( \rho_{t} + {\mathcal P}_{t})} \overline{\xi}^{\prime} ({\mathcal R} + \Psi)
+ \frac{\overline{\xi}\,{\mathcal H}}{a(\rho_{t} + {\mathcal P}_{t})} \Theta. 
\label{evolR3}
\end{eqnarray}
In Eq. (\ref{evolR}) and (\ref{evolR3}) we traded the difference of the Bardeen potentials for the total anisotropic stress as it follows 
directly from Eq. (\ref{diff}). 
Concerning Eqs. (\ref{evolR}) and (\ref{evolR3}) few comments are in order. We first note, as already mentioned, that the fluctuations 
of the viscous coefficients play the same r\^ole of the nonadiabatic pressure fluctuations $\delta p_{nad}$. In the 
 limit $\overline{\xi} \to 0$ the viscous coefficients do not contribute to the background but the fluctuations always
 affect the curvature perturbations. This is the situation leading to the viscous modes across matter-radiation equality 
 discussed in the first paper of Ref. \cite{mg1}. 

Equation (\ref{evolR}) leads to second-order equation for ${\mathcal R}$ which will be the basis for our subsequent considerations. 
By taking the time derivative of both sides of Eq. (\ref{evolR}) various terms will arise: the terms containing the Laplacian of $\Psi$ 
can will be eliminated through Eq. (\ref{evolR}) while the term proportional to the time derivative of the Laplacian of $\Psi$ (i.e. $\nabla^2 \Psi'$) can be expressed via the constraint (second equation of Eq. (\ref{G00G0i})) and in terms of Eq. (\ref{defRL}). 
After this straightforward algebraic procedure the explicit form of the second-order equation becomes: 
\begin{equation}
{\mathcal R}^{\prime\prime} + 2 \frac{z_{t}^{\prime}}{z_{t}} {\mathcal R}^{\prime} - c_{s}^2 \nabla^2 {\mathcal R} = 
\frac{ 3 a^4}{ z_{t}^2} \Pi + \Sigma_{{\mathcal R}}^{\prime} +  2 \frac{z_{t}^{\prime}}{z_{t}}\Sigma_{{\mathcal R}}, \qquad z_{t} = \frac{a^2 \sqrt{ \rho_{t} + {\mathcal P}_{t}}}{{\mathcal H} c_{s}}.
\label{evolR4}
\end{equation}
It is clear from the previous expressions that the presence of a gauge-invariant fluctuation of the 
bulk viscosity coefficient produces a computable source for the evolution of curvature perturbations which is 
fully equivalent to a nonadiabatic fluctuation of the pressure. 

As a cross-check it is useful to remark that, in the limit $\overline{\xi}\to 0$, $\Xi \to 0$ and $\Pi=0$, Eqs. (\ref{evolR3}) and (\ref{evolR4}) reproduce the well known results firstly obtained by Lukash \cite{lukash} in the absence of nonadiabatic pressure fluctuations (i.e. $\delta p_{nad} =0$). 
In this limit all the terms at the right hand side of Eq. (\ref{evolR4}) disappear and Eq. (\ref{evolR4}) becomes, as expected\footnote{In Eq. (\ref{evolR4a}) we have that 
 ${\mathcal P}_{t}$ coincides with $p_{t}$ since, in this case, the bulk viscosity coefficient vanishes. }
\begin{equation}
{\mathcal R}^{\prime\prime} + 2 \frac{z_{t}^{\prime}}{z_{t}} {\mathcal R}^{\prime} - c_{s}^2 \nabla^2 {\mathcal R} = 0,
 \qquad z_{t} = \frac{a^2 \sqrt{ \rho_{t} + p_{t}}}{{\mathcal H} c_{s}}.
 \label{evolR4a}
 \end{equation}
The case of a single scalar field is implicitly 
contained in Eq. (\ref{evolR4a}). Formally\footnote{From the purely algebraic viewpoint the situation is slightly more complicated. Indeed  the scalar field has an effective sound speed $c_{\varphi}^2 = 1 + 2 a^2 V_{, \varphi}/(3 {\mathcal H} \varphi^{\prime})$. In this case 
$\delta p_{nad} = \delta p_{\varphi} - c_{\varphi}^2 \delta \rho_{\varphi} = - 4 V_{,\varphi} \,\nabla^2\Psi/ ( 3 \ell_{P}^2 \varphi^{\prime} {\mathcal H})$ (where $V_{,\varphi} = \partial V/\partial\varphi$ and $V$ is the potential of the scalar field). These two modifications combine in 
Eq. (\ref{evolR4}) and lead to the standard form of the evolution equation for the normal modes.} the case of single scalar field is obtained by requiring $ c_{s}\to 1$, $\Sigma_{{\mathcal R}} \to 0$ and $\overline{\xi} =0$. In this case $(\rho_{t} + p_{t}) \to \varphi^{\prime\, 2}/a^2$ (where $\varphi$ denotes the scalar field) and $z_{t} \to z_{\varphi} = a \varphi^{\prime}/{\mathcal H}$. 

\subsection{Quasiadiabatic normal modes} 
Whenever $\Xi= \Xi(\rho_{t},\, H)$ the contribution of the bulk viscosity 
coefficient and of its fluctuations rearrange and cancel so that the source term is negligible 
at large scales. In spite of this observation the evolution equation for the curvature 
perturbations does not reproduce exactly the canonical result for the standard adiabatic modes, hence these solutions have been 
named quasiadiabatic. Since this is a particularly relevant point of the discussion we shall now present a more detailed analysis.

The term $\Sigma_{{\mathcal R}}$ appearing in Eq. (\ref{evolR}) contains two kinds of contributions: the terms 
that do not vanish in the limit $\xi \to 0$ and those that do vanish in the same limit.  By separating the terms of different origins Eq. (\ref{evolR}) can therefore be rewritten as 
 \begin{eqnarray}
{\mathcal R}' &=& \overline{\Sigma}_{{\mathcal R}} - \frac{2 {\cal H} c_{\rm s}^2}{\ell_{P}^2 \,a^2(\rho_{t} + {\mathcal P}_{t})}\nabla^2 \Psi
\nonumber\\
&+& \frac{3 {\mathcal H}^2 \, \Xi}{a (\rho_{t} + {\mathcal P}_{t})}   + \frac{ 3 {\mathcal H}}{a ( \rho_{t} + {\mathcal P}_{t})} \overline{\xi}'  ( {\mathcal R} + \Psi) + \frac{\overline{\xi} {\mathcal H}}{a ( \rho_{t} + {\mathcal P}_{t})} \Theta,
\label{evolR5}\\
\overline{\Sigma}_{{\mathcal R}} &=& - \frac{{\cal H}}{\rho_{t} + {\mathcal P}_{t}} \delta p_{\rm nad}  + \frac{{\mathcal H}}{(\rho_{t} + {\mathcal P}_{t})} \Pi.
\label{sigmaR}
\end{eqnarray}
We will now show that all the terms appearing in the second line at the right hand side o Eq. (\ref{evolR5}) are ${\mathcal O}(\nabla^2 {\mathcal R})$ and therefore negligible for typical length-scales larger than the Hubble radius. 

For this purpose let us notice immediately that the 
last term at the right hand side of Eq. (\ref{evolR5}) can also be written 
as  $\overline{\xi} \nabla^2({\mathcal R} + \Psi)/[a(\rho_{t} + {\mathcal P}_{t})] $. We shall therefore 
focus the attention on the first two terms appearing in the second line at the right hand side of Eq. (\ref{evolR5})
and remark  that in the linearized treatment $\Xi(\rho_{t},\, H)$ is equivalent to the case $\Xi(\rho_{t})$ since, according to the 
background equations, $H = \ell_{P} \sqrt{\rho_{t}/3}$. Since $\Xi$ defines the gauge-invariant fluctuation of the bulk viscosity 
given in Eq. (\ref{XI}) we can say, by definition, that 
\begin{equation}
\Xi = \biggl(\frac{\partial \overline{\xi}}{\partial \rho_{t}} \biggr)\delta \rho_{t} +  \overline{\xi}^{\prime} ( \beta - \alpha^{\prime}) = \biggl(\frac{\partial \overline{\xi}}{\partial \rho_{t}}\biggr) \delta \rho_{g},
\label{res1}
\end{equation}
where the second equality follows from the first one by recalling that $\overline{\xi}^{\prime} = (\partial \overline{\xi}/\partial \rho_{t})\rho_{t}^{\prime}$ and from the gauge invariant definition of $\delta \rho_{g}$ of Eq. (\ref{fluid}).  Thanks to Eq. (\ref{res1}) (and using the Hamiltonian constraint to express $\delta \rho_{g}$) we obtain:
\begin{eqnarray}
&& \frac{3 {\mathcal H}^2 \Xi}{a (\rho_{t} + {\mathcal P}_{t})} + \frac{ 3 {\mathcal H} \overline{\xi}^{\prime}}{a (\rho_{t} + {\mathcal P}_{t})} 
({\mathcal R} + \Psi) = 
\frac{3 {\mathcal H}}{a( \rho_{t} + {\mathcal P}_{t})} \biggl(\frac{\partial \overline{\xi}}{\partial \rho_{t}}\biggr) \biggl[ \frac{2 {\mathcal H}}{\ell_{P}^2 a^2 } \nabla^2 \Psi 
\nonumber\\
&& - \frac{6 {\mathcal H}^2 }{\ell_{P}^2 a^2} ( \Psi^{\prime} + {\mathcal H} \Phi) - 3 ( \rho_{t} + {\mathcal P}_{t}) 
({\mathcal R} + \Psi)\biggr].
\label{res2}
\end{eqnarray}
Using now Eq. (\ref{defRL}) we can immediately obtain, from Eq. (\ref{res2}) the following result
\begin{equation}
\frac{3 {\mathcal H}^2 \Xi}{a (\rho_{t} + {\mathcal P}_{t})} + \frac{ 3 {\mathcal H} \overline{\xi}^{\prime}}{a (\rho_{t} + {\mathcal P}_{t})} 
({\mathcal R} + \Psi)  = - \frac{ 2 {\mathcal H}}{a^3 \ell_{P}^2 (\rho_{t} + {\mathcal P}_{t})^2 } \overline{\xi}^{\prime} \nabla^2 \Psi.
\label{res3}
\end{equation}
Inserting now Eq. (\ref{res3}) into Eq. (\ref{evolR5}) and recalling the momentum constraint to eliminate $\Theta$ we finally arrive at the 
equation
\begin{equation}
{\mathcal R}^{\prime} = \overline{\Sigma}_{{\mathcal R}} 
+ \frac{\overline{\xi} \nabla^2 {\mathcal R}}{a ( \rho_{t} + {\mathcal P}_{t})} - \frac{{\mathcal H}\, \, \nabla^2 \Psi }{4 \pi G a^2 ( \rho_{t} + {\mathcal P}_{t})}\biggl( c_{s}^2 + \frac{\overline{\xi}^{\prime}}{a ( \rho_{t} + {\mathcal P}_{t})} - \frac{\overline{\xi} \ell_{P}^2 a}{2 {\mathcal H}}\biggr), 
\label{res5}
\end{equation}
where all the terms containing the Laplacian of $\Psi$ have been collected: the quantity appearing in the round bracket is actually  ${\mathcal P}_{t}^{\prime}/\rho_{t}^{\prime}$ so that 
Eq. (\ref{res5}) becomes:
\begin{equation}
{\mathcal R}^{\prime} = \overline{\Sigma}_{{\mathcal R}} 
+ \frac{\overline{\xi} \nabla^2 {\mathcal R}}{a ( \rho_{t} + {\mathcal P}_{t})} - \frac{{\mathcal H}\, c_{eff}^2\,\nabla^2 \Psi }{4 \pi G a^2 ( \rho_{t} + {\mathcal P}_{t})},
\label{res6}
\end{equation}
where $c_{eff}^2 = {\mathcal P}_{t}^{\prime}/\rho_{t}^{\prime}$ is just an auxiliary variable\footnote{ It must be stressed that 
$c_{eff}^2$ does have the physical meaning of a sound speed only in the limit $\overline{\xi} \to 0$ since, in this limit, 
$c_{eff}^2 $ coincides with $c_{s}^2$. Whenever $\overline{\xi} \neq 0$ the effective sound speed is not
only given by $c_{eff}^2$, as we shall demonstrate in a moment.}. 
From Eq. (\ref{res6})  the second-order form of the evolution equation of ${\mathcal R}$ can be written as:
\begin{eqnarray}
 {\mathcal R}^{\prime\prime} + 2 \frac{\overline{z}_{t}}{\overline{z}_{t}} {\mathcal R}^{\prime} - c_{eff}^2 \nabla^2 {\mathcal R} = \biggl[ \frac{\overline{\xi}\, \nabla^2 {\mathcal R}^{\prime}}{a ( \rho_{t} + {\mathcal P}_{t})}\biggr] 
 + 2 \frac{\overline{z}_{t}^{\prime}}{\overline{z}_{t}} \frac{\overline{\xi} \, \nabla^2 {\mathcal R}}{a ( \rho_{t} + {\mathcal P}_{t})} , \qquad 
\overline{z}_{t} = \frac{a^2 \sqrt{\rho_{t} + {\mathcal P}_{t}}}{{\mathcal H} |c_{eff}|}.
\label{res7}
\end{eqnarray}
In the case $\delta p_{nad} \to 0$ (absence of nonadiabatic pressure fluctuations) and  $\Pi\to 0$ (absence of anisotropic stress) we also have $\overline{\Sigma}_{{\mathcal R}} \to 0$. When the preceding conditions are all met in the limot $\overline{\xi} \to 0$, the result of Eq. (\ref{res7}) reproduces the results of Lukash \cite{lukash} since, in this case, $c_{eff}^2$ coincides with $c_{s}^2$. Whenever $\overline{\xi} \neq 0$, however, the situation is totally different and Eq. (\ref{res7}) describes, as anticipated, the evolution of the quasiadiabatic modes.

\subsection{Evolution of the quasiadiabatic normal mode}

To analyze the evolution of the quasiadiabatic modes it is natural to set to zero both the anisotropic stress and the
nonadiabatic pressure fluctuations. Going to Fourier space Eq. (\ref{res7}) can also be written as:
\begin{equation}
{\mathcal R}^{\prime\prime} 
+ \biggl[ 2 \frac{\overline{z}_{t}^{\prime}}{\overline{z}_{t}} + \frac{k^2 \overline{\xi}}{a( \rho_{t} + {\mathcal P}_{t})} \biggr] {\mathcal R}^{\prime}  
+ k^2 \biggl\{ c_{eff}^2 + \biggl[ \frac{\overline{\xi}}{a ( \rho_{t} + {\mathcal P}_{t})}\biggr]^{\prime} + 2 \frac{\overline{z}_{t}^{\prime}}{\overline{z}_{t}}
 \frac{\overline{\xi}}{a( \rho_{t} + {\mathcal P}_{t})} \biggr\} {\mathcal R} =0.
 \label{res8}
\end{equation}
Equation (\ref{res8}) is homogeneous but it is non-standard insofar as  the pump fields get corrected both inside and (partially) outside the Hubble radius.  We shall now argue that the coefficient of the third term of Eq. (\ref{res8}) 
is positive semi-definite if the inflationary phase is driven by the bulk viscosity coefficient. To demonstrate 
this point it is useful to reverse the question and demand that the coefficient of the third term in Eq. (\ref{res8}) is positive 
semi-definite; such a request implies:
\begin{equation}
c_{eff}^2 + \frac{1}{\overline{z}_{t}^2} \frac{\partial}{\partial \tau} \biggl[ \frac{a^3 \overline{\xi}}{{\mathcal H} c_{eff}^2} \biggr] \geq 0.
\label{ineq1}
\end{equation}
Assuming now that the inflationary phase is triggered by a dynamical bulk viscosity coefficient we will have 
\begin{equation}
\dot{H} = - \frac{3}{2} ( 1 + w) H^2 + \frac{3}{2} \ell_{P}^2 H \overline{\xi}.
\label{res10}
\end{equation}
 From Eq. (\ref{res10}) we can argue that $\overline{\xi}$ is always expressible as
 \begin{equation}
 \overline{\xi} = H \, \overline{M}^2_{P} (1 + w) \biggl[ 1 - \frac{2 \epsilon}{3 ( 1 + w)} \biggr]
 \label{res1a}
 \end{equation}
 where $\epsilon = - \dot{H}/H^2$ is the standard slow-roll parameter. During inflation $\epsilon \ll 1$ and therefore 
 we shall have, in the first approximation,  that $ \overline{\xi} \simeq H \, \overline{M}^2_{P} (1 + w)$. 
 The inequality of Eq. (\ref{ineq1}) becomes immediately 
\begin{equation}
 c_{eff}^2 + \frac{c_{eff}^2 H^2}{a ( \rho_{t} + {\mathcal P}_{t})} \frac{\partial}{\partial t} \biggl(\frac{a \overline{\xi} }{ H^2 c_{eff}^2} \biggr) \geq 0.
\end{equation}
 Using now Eq. (\ref{res1}) and the fact that $c_{eff}^2$ is asymptotically constant during inflation it is possible to prove that Eq. (\ref{ineq1}) implies that\footnote{As it ca ne explicitly verified from its definition and from the asymptotic expression of $\overline{\xi}$ we have 
 that $c_{eff}^2 \to c_{s}^2 - \gamma - 1$ up to slow-roll corrections which are subleading.} 
$ \epsilon \leq (1 + w)/(1 + w - 2 c_{s}^2)$: all the terms at the right hand side of the previous inequality are ${\mathcal O}(1)$. Therefore 
the inequality simply requires that $\epsilon < 1$ which is always true since, by definition,  $\epsilon \ll 1 $
during the slow-roll phase.

The term containing the first derivative of ${\mathcal R}$ can be eliminated and therefore Eq. (\ref{res8}) 
 can be written, in Fourier space, as
\begin{eqnarray}
&& q^{\prime\prime} + \biggl[ k^2 {\mathcal C}^2(k,\tau) - \frac{\overline{z}_{t}^{\prime\prime}}{\overline{z}_{t}} \biggr] q =0,
\nonumber\\
&& {\mathcal C}^2(k,\tau) = c_{eff}^2 + \frac{1}{ 2 \overline{z}_{t}^2} \biggl[ \frac{\overline{z}_{t}^2 \overline{\xi}}{a ( \rho_{t} + {\mathcal P}_{t})} \biggr]^{\prime}
- \frac{ k^2 \overline{\xi}^2}{4 a^2 ( \rho_{t} + {\mathcal P}_{t})^2},
\label{MODEF1}
\end{eqnarray}
where the variable $q$ is implicitly defined as 
\begin{equation}
 \qquad W {\mathcal R} = q,\qquad \frac{W^{\prime}}{W} = \frac{\overline{z}_{t}^{\prime}}{\overline{z}_{t}} + \frac{k^2 \overline{\xi}}{2 a (\rho_{t} + {\mathcal P}_{t})}.
 \end{equation}
If  $\xi$ is assigned with the constraint that asymptotically there is a slow-roll phase of quasi de-Sitter type we must always demand,
\begin{equation}
\epsilon = - \frac{\dot{H}}{H^2}, \qquad \frac{\partial_{t} \overline{\xi}}{ H \overline{\xi}}  \ll 1, \qquad 
\frac{ \partial_{t} (a \sqrt{\rho_{t} + {\mathcal P}_{t})}}{H a \sqrt{\rho_{t} + {\mathcal P}_{t}}} \ll 1.
\label{SR}
\end{equation}
The solutions pinned down by the conditions (\ref{SR}) are clearly not the most general 
ones compatible with the condition that the bulk viscosity coefficient depends solely on the energy 
density of the plasma. At the same time these solutions will be useful for a specific 
comparison of the quasiadiabatic mode with the genuine adiabatic solution to be discussed 
in section \ref{sec5}. Note in particular that there are solutions of Eqs. (\ref{FL}) where 
the slow-roll parameters of Eq. (\ref{SR}) are time-independent\footnote{These solutions are well known (see e.g. \cite{bv1}). In the 
power-law case a good example is the standard solution $a(t) \propto (t/t_{1})^{\alpha}$ with $\alpha \gg 1$ and $\overline{\xi} \propto 
H \overline{M}_{P}^2$. In this case, for instance,  $\epsilon \to 1/\alpha$. }. Even if the equations 
for the quasiadiabatic modes are completely general we shall find it convenient, for the 
sake of simplicity, to assume that the slow-roll parameters are constant at least approximately.
Since to leading order in the slow-roll parameters  $\overline{\xi} \propto H \overline{M}_{P}^2$
the explicit form of ${\mathcal C}(k \tau)$ becomes
\begin{equation}
{\mathcal C}^2(k,\tau) = c_{s}^2 - \frac{3}{4}(1 + w) + \frac{1 + w}{4 \epsilon} - \frac{k^2 \tau^2 ( 1 + w)^2 }{16 \epsilon^2}.
\end{equation}
It is clear that during the inflationary expansion (or during the fully developed inflationary phase) 
the numerical value of ${\mathcal C}^2(k\tau)$ is dominated by the third and the fourth terms of the previous expression.
The equation to be solved becomes, therefore, 
\begin{equation}
q^{\prime\prime} +  \biggl[ - \frac{(1 + w)^2}{16 \epsilon^2} k^4 \tau^2 + \frac{(1 + w) k^2}{4 \epsilon}  - \frac{\overline{z}_{t}^{\prime\prime}}{\overline{z}_{t}} \biggr] \,\,q =0.
\label{sol0}
\end{equation}
This equation can be studied in two opposite regimes, namely
\begin{eqnarray}
&& q^{\prime\prime} - \frac{\overline{z}^{\prime\prime}_{t}}{\overline{z}_{t}} \,\,q=0 , \qquad \frac{k^2}{\epsilon} \ll \frac{\overline{z}_{t}^{\prime\prime}}{\overline{z}_{t}},
\label{sol1}\\
&& q^{\prime\prime} + \biggl[ - \frac{(1 + w)^2}{16 \epsilon^2} k^4 \tau^2 + \frac{(1 + w) k^2}{4 \epsilon} \biggl] \,\,q =0, \qquad  
\frac{k^2}{\epsilon} \gg \frac{z_{t}^{\prime\prime}}{z_{t}}.
\label{sol2}
\end{eqnarray}
The solution of Eq. (\ref{sol1}) is standard and describes the regime when the relevant wavelengths are larger 
than the Hubble radius: 
\begin{equation}
q(k,\tau) = A_{k} \overline{z}_{t}(\tau) + B_{k}  \overline{z}_{t}(\tau) \int^{\tau} \frac{d\tau^{\prime}}{\overline{z}^2_{t}(\tau^{\prime})}.
\label{sol1a}
\end{equation}
 For a proper normalization of the whole solution 
(and for a correct determination of the scalar power spectrum) we need also to solve Eq. (\ref{sol2}) and, in this case, the solution 
can be solved in terms of parabolic cylinder functions. In the case of Eq. (\ref{sol2}), however, the equation gets even simpler:
\begin{equation}
\frac{ d^2 q}{d x^2} + [ b - b^2 x^2] q=0, \qquad  x = k \tau, \qquad b = \frac{1+ w}{4 \epsilon}, 
\end{equation}
whose solution is given by 
\begin{equation}
q(k,\tau) = C_{1}(k) e^{-\frac{(1 + w)}{8\epsilon} k^2 \tau^2} + C_{2}(k) e^{\frac{(1 + w)}{8\epsilon} k^2 \tau^2}.
\label{sol2a}
\end{equation}
The solution parametrized by the coefficient $C_{2}(k)$ is physically unacceptable. Taking into account that 
$k \tau >1$ and that $\epsilon < 1$ this solution describes the situation where starting from perturbatively small initial data 
the curvature perturbations quickly become ${\mathcal O}(1)$ and rapidly jeopardize perturbation theory. 
We are therefore left with the  solution parametrized by $C_{1}(k)$ describing a the exponential suppression of the curvature 
perturbations in the regime $k \tau > 1$. Up to a phase we shall normalize $|C_{1}(k)|= 1/\sqrt{2 k}$. This choice will be useful
when comparing the quasiadiabatic power spectra with their adiabatic counterpart. 

\renewcommand{\theequation}{4.\arabic{equation}}
\setcounter{equation}{0}
\section{Nonlinear curvature perturbations}
\label{sec4}
\subsection{ADM decomposition and normal coordinates}
In section \ref{sec3} we have shown that the gauge-invariant inhomogeneities 
of the viscous coefficients provide a supplementary source term in the evolution equations of curvature 
perturbations.  If the bulk viscosity coefficient only depends on the energy density and on the Hubble rate the resulting 
curvature perturbations are approximately conserved over large-scales even if the evolution of the corresponding normal modes 
differs substantially from the perfect fluid case. It is desirable to scrutinize the validity and the implications of this result without relying on the perturbative expansion. To achieve this goal we shall study the effects of the bulk and shear viscosity within the expansion in spatial gradients. This
technique  has been used for the first time in the analysis of general relativistic singularities \cite{bel2} and subsequently 
exploited in a variety of contexts ranging from inflationary models \cite{star,tomita} to large-scale structure \cite{NH}. 

In the Arnowitt-Deser-Misner formalism \cite{ADM1,shap} (ADM in what follows) the line element is expressed in terms of the 
conventional $(3+1)$-dimensional decomposition:
\begin{equation}
ds^2 = g_{\mu\nu}(\tau,\vec{x})\, dx^{\mu} \, dx^{\nu} =N^2 d\tau^2 - \gamma_{ij} (d x^{i} + N^{i} d\tau)   ( d x^{j} + N^{j} d\tau),
\label{ad1}
\end{equation}
where $N=N(\tau,\vec{x})$ denotes the lapse function, $N^{i}=N^{i}(\tau,\vec{x})$ is the shift vector and $\gamma_{ij}=\gamma_{ij}(\tau, \vec{x})$ is the three-dimensional metric.  In the ADM variables of Eq. (\ref{ad1}) the extrinsic curvature of the spatial slices will be denoted by 
$K_{ij}(\tau,\vec{x})$ and it is defined as $K_{ij}(\tau,\vec{x}) =  [- \partial_{\tau}\gamma_{ij} + \nabla_{i}N_{j} + \nabla_{j} N_{i}]/(2 N)$ where $\nabla_{i}$ are the covariant derivatives defined with respect to $\gamma_{ij}$.
The other standard notations for the traces are $K= \gamma^{ij} K_{ij}$ and $\mathrm{Tr}K^2 = K_{i}^{j} \, K_{j}^{i}$. The traceless part of the extrinsic curvature will be denoted by $\overline{K}_{i}^{j} = K_{i}^{j} - \delta_{i}^{j} K/3$.  For the sake of simplicity the shift vectors will be 
assumed to vanish (i.e. $N^{i}=0$) and in this case the coordinate observers coincide 
with the normal observers\footnote{This is a choice often made in numerical relativity \cite{shap} when imposing Gaussian 
normal coordinates. In the present case, however, we shall not use literally the normal coordinates since we shall keep 
the lapse function generic with the purpose of making specific contact with the linearized treatment of the fluctuations.}. 
 
\subsection{Nonlinear evolution, curvature modes and viscosity}
In the viscous case the nonlinear generalization of the curvature perturbations on comoving orthogonal hypersurfaces is:
\begin{equation}
{\mathcal R}_{i}(\tau,\vec{x}) = \frac{1}{3} \nabla_{i} [\ln{(\sqrt{\gamma})}] - \frac{1}{3N} \partial_{\tau}[ \ln{(\sqrt{\gamma})}] \,\, u_{i}, 
\label{defRNL}
\end{equation}
where $\gamma = {\mathrm det} [ \gamma_{ij}]$ and $u_{i}$ is the spatial component of $u_{\mu}$. The nonlinear generalization  of the density contrast on uniform curvature hypersurfaces becomes instead
\begin{equation}
\zeta_{i}(\tau,\vec{x}) =  \frac{1}{3} \nabla_{i}[ \ln{(\sqrt{\gamma})} ]+ \frac{\nabla_{i} \rho}{ 3(\rho + {\mathcal P})},
\label{defZNL}
\end{equation}
where $\rho$ and ${\mathcal P}= p + 3 \xi K$ are now nonlinear quantities; the 
expression of ${\mathcal P}$ and $\rho$ holds to lowest order in the gradient expansion {\em but are not necessarily homogeneous} and this is 
why we distinguished them from their corresponding background values denoted, respectively,  by ${\mathcal P}_{t}$ and 
$\rho_{t}$ in sections \ref{sec3} and \ref{sec4}.

Equations (\ref{defRNL}) and (\ref{defZNL}) define a set of nonlinear variables which are 
also gauge-invariant. Both ${\mathcal R}_{i}$ and $\zeta_{i}$ do not depend on the choice of time hypersurfaces and are exactly invariant for infinitesimal coordinate transformations in the perturbative regime.  Equations (\ref{defRNL}) and (\ref{defZNL}) correspond, in linear theory, to the variables ${\mathcal R}$ and $\zeta$. Indeed in the conformally Newtonian frame where the gauge freedom is removed and the coordinate 
system is completely fixed and $N^2(\tau,\vec{x}) = a^2(\tau) [ 1 + 2 \phi(\tau,\vec{x})]$ and $w_{i j}(\tau,\vec{x}) = a^2(\tau)[ 1 - 2 \psi(\tau,\vec{x})]\delta_{ij}$. In the limit set by the two preceding expressions we have that ${\mathcal R}_{i} \to \partial_{i} {\mathcal R}$ and $\zeta_{i} \to   \partial_{i} \zeta$ where ${\mathcal R} = [ - \psi - {\mathcal H} ( \psi^{\prime} + {\mathcal H} \phi)/({\mathcal H}^2 - {\mathcal H}^{\prime})]$ and $\zeta = - \psi 
+ \delta\rho_{t} /[ 3 (\rho_{t} + {\mathcal P}_{t})]$ and coincide therefore with the expressions of Eqs. (\ref{defRL}) and (\ref{defzetaL}). In the linearized 
approximation the Eqs. (\ref{defRNL}) and (\ref{defZNL}) are invariant under {\em infinitesimal} coordinate transformations. 
In the general case they are also invariant under {\em finite} coordinate transformations and that preserve the order of the gradient 
expansion. These transformations are of the type $\tau \to T = T(\tau, \vec{x})$ and $x^{k} \to X^{k}(\tau, \vec{x}) = f^{k}(\tau, \vec{x}) + F^{k}(\tau, \vec{x})$ where $F^{k}$ contains at least one spatial gradient\footnote{ We shall be working to lowest order in the gradient expansion which means, in particular, that the trace of the extrinsic curvature, the energy density, the pressure and the viscous coefficients will all be 
fully inhomogeneous but will not contain any spatial gradient while ${\mathcal R}_{i}$ and $\zeta_{i}$ will contain at most one spatial gradient. 
This means, in the present case, that $F^{k}$ will contain only one spatial gradient.}. Probably the first nonlinear generalization of inflationary curvature perturbations has been proposed in \cite{salope} after the pioneering analyses on the gauge-invariant treatment of linearised cosmological perturbations \cite{bard}.  Similar variables have been subsequently scrutinized and rediscovered by different authors \cite{shell}. 

Although the nonlinear evolution of the curvature perturbations caused by the inhomogeneities in the viscous coefficients can be followed either in $\zeta_{i}$  or in ${\mathcal R}_{i}$, it appears to be more useful in the latter than in the former since  
${\mathcal R}$ is directly related to the normal mode of the system. In the presence of viscous stresses the nonlinear evolution 
of ${\mathcal R}_{i}$ is given by: 
\begin{equation}
\partial_{\tau} {\mathcal R}_{i} = \frac{1}{3} \partial_{\tau} \biggl(\frac{\partial_{i}\rho}{\rho + {\mathcal P}}\biggr) - \frac{1}{3} \partial_{i} \biggl(\frac{\partial_{\tau}\rho}{\rho + {\mathcal P}}\biggr) + \, .\,.\,.\,
\label{evR1}
\end{equation}
where the ellipses stand for terms which contain, at least, three spatial gradients  and are therefore of higher order in the gradient expansion.
Equation (\ref{evR1}) generalizes the results of Refs. \cite{salope,shell} (see also \cite{mg1} second paper). 
For the sake of conciseness the ellipses shall be neglected altogether in the subsequent discussions but it is  understood 
that the forthcoming results hold to lowest order in the gradient expansion.
Recalling the expression of the effective pressure the right hand side of Eq. (\ref{evR1}) can be made more explicit; the result is:
\begin{equation} 
\partial_{\tau} {\mathcal R}_{i} = \frac{[\partial_{\tau} \rho \,\,\partial_{i} p - \partial_{i} \rho \,\,\partial_{\tau} p] + [\partial_{\tau} \rho\,\, \partial_{i} ( K \, \xi) - \partial_{i} \rho \,\,\partial_{\tau}(K \xi)]  }{3( \rho + {\mathcal P})^2},
\label{evR2}
\end{equation}
where the terms have been grouped in such a way that each of the two square brackets reproduces, respectively, the nonadiabatic and 
the viscous contributions in the perturbative limit.  To further simplify the right hand side of Eq. (\ref{evR2}) we can use first 
the evolution of $\rho$ (i.e. $\partial_{\tau} \rho = K\, N\, (\rho + {\mathcal P})$) and then rearrange the various terms. The result of this 
step is given by:
\begin{eqnarray}
\partial_{\tau} {\mathcal R}_{i} &=& \frac{ K N}{ 3 (\rho + {\mathcal P})} \bigl( \partial_{i} p - c_{s}^2 \partial_{i} \rho\bigr)
\nonumber\\
&+& \frac{\xi}{3 (\rho + {\mathcal P})^2} \bigl[(\partial_{\tau} \rho)\partial_{i} K - (\partial_{i} \rho)\partial_{\tau} K \bigr]
\nonumber\\
&+& \frac{K}{3 ( \rho + {\mathcal P})^2} \bigl[ (\partial_{\tau} \rho) \partial_{i} \xi - \partial_{i} \rho (\partial_{\tau} \xi)\bigr].
\label{evR3}
\end{eqnarray}
The first at the right hand side of Eq. (\ref{evR3}) is  the nonadiabatic pressure fluctuation;  the second term at the right hand side of Eq. (\ref{evR3}) vanishes since its contribution is of higher order in the gradients. More specifically this term can be rewritten as:
\begin{equation}
\frac{\xi}{3 (\rho + {\mathcal P})^2} \bigl[(\partial_{\tau} \rho)\partial_{i} K - 
(\partial_{i} \rho)\partial_{\tau} K \bigr] = \frac{\xi N}{6 (\rho+ {\mathcal P})} \partial_{i}\bigl( K^2 - 3 \ell_{P}^2 \rho),
\label{evR4}
\end{equation}
but the term at the right hand side vanishes. Indeed the inhomogeneous Eintsein equations are  
\begin{equation}
2 \ell_{P}^2 \rho = K^2 - \mathrm{Tr}K^2,\qquad 3 N \ell_{P}^2 (\rho + {\mathcal P}) = 2 \partial_{\tau}K - 3 N \mathrm{Tr}K^2 + N K^2.
\label{evR5}
\end{equation}
In the first equation 
$\mathrm{Tr}K^2 = K^2/9 + {\mathrm Tr} \overline{K}^2$ and ${\mathrm Tr}\overline{K}^2$ is of higher order\footnote{This statement will be 
specifically demonstrated in the last part of this section.}. 
In summary, thanks to the results of Eq. (\ref{evR5}) and in the absence of nonadiabatic pressure fluctuations,    Eq. (\ref{evR3}) becomes:
\begin{equation}
 \partial_{\tau} {\mathcal R}_{i} = \frac{K}{3 ( \rho + {\mathcal P})^2} 
 \bigl[ (\partial_{\tau} \rho) \partial_{i} \xi - \partial_{i} \rho (\partial_{\tau} \xi)\bigr]. 
\label{evR6}
\end{equation} 
If the source term in Eq. (\ref{evR6}) vanishes the curvature inhomogeneities will be conserved and the equations of motion will enjoy a further 
symmetry\footnote{According to the results obtained so far,
Eq. (\ref{evR1}) is invariant for ${\mathcal R}_{i}(\tau, \vec{x}) \to {\mathcal R}_{i}(\tau, \vec{x}) + {\mathcal Q}(\vec{x})$ provided
$\xi$ is either a space-time constant or a function of the total energy density.} since ${\mathcal R}_{i}(\tau,\vec{x})$ can be shifted by a a term constant in time (but not in space). 
When $\xi = \xi(\rho)$ Eq. (\ref{evR6}) implies that  $\partial_{\tau} {\mathcal R}_{i} =0$: in this case the two terms 
at the right hand side simplify because $\partial_{i} \xi = (\partial \xi/\partial \rho)\partial_{i} \rho$ and $\partial_{\tau}\xi 
= (\partial \xi/\partial \rho)\partial_{\tau} \rho$.   

\subsection{Bulk viscosity versus shear viscosity}

To obtain Eqs. (\ref{evR4}), (\ref{evR5}) and (\ref{evR6}) two results have been used, namely that $\mathrm{Tr}\overline{K}^2$ is of higher order in the gradient expansion and that the shear viscosity does not contribute to the evolution to leading order in the gradient expansion. These two 
results also imply  that bulk viscosity does contribute to the deceleration parameter while the contribution of the shear 
viscosity is of higher order. Given their relevance for the nonlinear discussion, these two points will now be discussed in some detail.
For the purposes of this discussion we shall write explicitly the Einstein equations in their contracted form\footnote{In other 
words we shall give, respectively, the $(00)$, $(0i)$ and $(ij)$ components of the equations written in the form 
$R_{\mu}^{\nu} = \ell_{P}^2 [{\mathcal T}_{\mu}^{\nu} - \delta_{\mu}^{\nu} \, {\mathcal T}/2]$.}
\begin{eqnarray}
&& \frac{1}{N}\partial_{\tau} K - \mathrm{Tr} K^2 = \frac{\ell_{P}^2}{2}  ( \rho + 3 {\mathcal P}),
\label{Ei1}\\
&& \nabla_{i} K - \nabla_{k} K^{k}_{i} = N \ell_{P}^2 \biggl[ ( \rho + {\mathcal P}) u_{i} u^{0} + 2\eta \overline{K}_{i}^{j} u_{j} u^{0} \biggr],
\label{Ei2}\\
&& \frac{1}{N}\partial_{\tau} K_{i}^{j} -  K K_{i}^{j} -  r_{i}^{j} = \ell_{P}^2 \biggl[ \frac{({\mathcal P} - \rho)}{2} \delta_{i}^{j} - 2 \eta \overline{K}_{i}^{j} + \Pi_{i}^{j} \biggr],
\label{Ei3}
\end{eqnarray}
where $\Pi_{i}^{j}$ denotes the  anisotropic stress (which is by definition a traceless rank-two tensor in three dimensions) and 
$r_{ij}$ are the components of the Ricci tensor of the spatial slices\footnote{By definition 
$r_{ij}(\tau,\vec{x}) = \partial_{m} \, ^{(3)}\Gamma^{m}_{ij} -\partial_{j} ^{(3)}\Gamma_{i m}^{m} + ^{(3)}\Gamma_{i j}^{m} 
\,^{(3)}\Gamma_{m n}^{n} - ^{(3)}\Gamma_{j n}^{m} \,^{(3)}\Gamma_{i m}^{n}$ where the Christoffel connections 
are computed in terms of $\gamma_{ij}$.}. Both $\Pi_{i}^{j}$ and $r_{i}^{j}$ are of higher order in the gradients and we shall see that the traceless part of the extrinsic curvature is also of higher order in the gradients and it is the only component affected by the presence of shear viscosity.
Indeed, after taking the  the traceless part of Eq. (\ref{Ei3}) the following equation is obtained:
\begin{equation} 
\partial_{\tau} \overline{K}_{i}^{j} - N K \overline{K}_{i}^{j}  = - 2 \eta N \ell_{P}^2  \overline{K}_{i}^{j} + N \ell_{P}^2 \Pi_{i}^{j} + N \overline{r}_{i}^{j},
\label{S1}
\end{equation}
where $\overline{r}_{i}^{j} = (r_{i}^{j} - \delta_{i}^{j} \, r/3)$ is traceless. Equation (\ref{S1}) shows that the shear 
viscosity (unlike bulk viscosity) completely decouples from the trace of the extrinsic curvature and only 
affects the traceless part.  When $\eta(\tau, \vec{x})$ Eq. (\ref{S1}) can be easily solved and the result is\footnote{Recall that 
$N\, K = - (\partial_{\tau} \sqrt{\gamma})  /\sqrt{\gamma}$.}
\begin{eqnarray}
\overline{K}_{i}^{j}(\tau,\vec{x}) &=&  \frac{\ell_{P}^2}{\sqrt{\gamma(\tau,\vec{x})}} \int_{\tau_{*}}^{\tau} d\tau^{\prime\prime} \sqrt{\gamma(\tau^{\prime\prime},\vec{x})}
N(\tau^{\prime\prime},\vec{x})\, e^{ - 2 {\mathcal A}(\tau^{\prime\prime},\tau,\vec{x})}\, \Pi_{i}^{j}(\tau^{\prime\prime}, \vec{x}) d\tau^{\prime\prime}
\nonumber\\
&+& \frac{1}{\sqrt{\gamma(\tau,\vec{x})}} \int_{\tau_{*}}^{\tau} d\tau^{\prime\prime} \sqrt{\gamma(\tau^{\prime\prime},\vec{x})}
N(\tau^{\prime\prime},\vec{x})\, e^{ - 2 {\mathcal A}(\tau^{\prime\prime},\tau,\vec{x})}\, \overline{r}_{i}^{j}(\tau^{\prime\prime}, \vec{x}) d\tau^{\prime\prime}
\nonumber\\
&+& \frac{\sqrt{\gamma(\tau_{*}, \vec{x})}}{\sqrt{\gamma(\tau,\vec{x})}} \, \overline{K}_{i}^{j}(\tau_{*},\vec{x}) e^{- 2 {\mathcal A}(\tau_{*}, \tau, \vec{x})}, 
\nonumber\\
 {\mathcal A}(\tau_{1},\tau_{2},\vec{x}) &=& \ell_{P}^2 \int_{\tau_1}^{\tau_{2}} \eta(\tau^{\prime},\vec{x}) \, N(\tau^{\prime},\vec{x}) \, d\tau^{\prime},
\label{S2}
\end{eqnarray}
where $\tau_{*}= \tau_{*}(\vec{x})$ denotes some arbitrary integration time while, in the last line, $\tau_{1}$ and $\tau_{2}$ denote
 two generic times. Equation (\ref{S2})  demonstrates that the traceless part of the extrinsic curvature 
is determined by the anisotropic stress and by the traceless part of the intrinsic curvature. Both 
quantities are of higher order in the gradient expansion. The last term at the right hand side of 
Eq. (\ref{S2}) shows that the shear viscosity suppresses the traceless part of the extrinsic curvature even further in comparison with the case $\eta \to 0$. 

Since the evolution of $\eta$ decouples from the trace of the extrinsic curvature, it does not 
contribute to the inhomogeneous generalization of the deceleration parameter: the only accelerated 
solutions obtainable in the case of irreversible fluids are determined by the bulk viscosity coefficient. For the sake of comparison with the fully homogenous case we choose Gaussian normal coordinates and set $N=1$; in this situation Eq. (\ref{Ei1}) can be written as:
\begin{eqnarray}
q(t,\vec{x}) \mathrm{Tr} K^2 &=& \ell_{P}^2 \biggl[ (\rho + {\mathcal P}) u_{0} u^{0} + \frac{{\mathcal P}- \rho}{2}\biggr] 
\nonumber\\
&+& 2 \eta\biggl[ u^{k} u^{\ell} \sqrt{1 + u^2} \overline{K}_{k \ell} - \frac{2}{3} u^2 \partial_{t} \sqrt{1 + u^2}
\nonumber\\
 &+& 
u^{k} \sqrt{1 + u^2} \partial_{k} \sqrt{1 + u^2} - \frac{u^2}{3 \sqrt{\gamma}} \partial_{k} ( \sqrt{\gamma} u^{k})\biggr].
\label{00cont}
\end{eqnarray}
where $q(\vec{x},t) = -1 + \dot{K}/{\rm Tr} K^2$ is the inhomogeneous generalization of the deceleration parameter\footnote{In the homogeneous and isotropic limit (i.e. $ \gamma_{ij} = a^2(t) 
\delta_{ij}$)  $q(t) \to - \ddot{a} a/ \dot{a}^2$ where the overdot denotes the derivative with 
respect to the cosmic time coordinate $t$ which coincides with $\tau$ in the case Gaussian normal coordinates (i.e. $N=1$ and $N^{i} =0$). }.
To discuss the sign of $q(t, \vec{x})$ the following three remarks are in order: {\it i)}  since $\gamma^{ij}$ is always positive semi-definite we have that 
 $u_{0}\,u^{0} = 1 + u^2  \geq 1$ (where, we remind, $u^2= \gamma^{ij} u_{i} u_{j}$);
{\it ii)} with the preceding observation it follows from the first two terms at the right hand side of Eq. (\ref{00cont}) 
that $q(t,\vec{x})$ is always positive semi-definite as long as $(\rho + 3 {\mathcal P}) \geq 0$ up to correction ${\mathcal O}(u^2)$; {\it iii)} 
the terms multiplying the shear viscosity $\eta$ all contain at least to gradients since each velocity field carries at least one 
gradient because of the momentum constraint of Eq. (\ref{Ei2}).  
 
Thanks to the three previous observations the sign of the generalized 
deceleration parameter only depends on ${\mathcal P} = p - \xi \nabla_{\alpha}u^{\alpha} \simeq p + K \xi$ (and hence on the bulk viscosity
coefficient) while the shear viscosity does not play any r\^ole. According to Eq. (\ref{00cont}) bulk viscosity only enters  to second order in the gradient expansion where, however, the bulk viscosity also contributes through the term $(\rho + {\mathcal P}) u^2$ implicitly contained in $ (\rho + {\mathcal P}) u_{0} u^{0}$. In conclusion the proof presented in the first part of this section is now complete and curvature perturbations are nonlinearly  
conserved if and when the bulk viscosity coefficient only depend on the energy density (or on the trace of the extrinsic curvature).  

\renewcommand{\theequation}{5.\arabic{equation}}
\setcounter{equation}{0}
\section{Tensor to scalar ratio}
\label{sec5}
After the scrutiny of the nature of the quasiadiabatic modes both at the 
linear and at the nonlinear level, it seems useful to draw an explicit comparison between the quasiadiabatic solution and the genuine adiabatic 
paradigm. In this respect the simplest and most revealing  quantity to estimate is the amplitude of the tensor to scalar 
ratio, i.e. the ratio between the scalar and tensor power spectra. As in section \ref{sec3} we shall assume a
slow-roll phase is only supported by the bulk viscosity (see, in particular, Eq. (\ref{SR}) and discussions therein). 
This case will then be compared with the conventional situation of a single field inflationary model. Even if the discussion could be conducted in fairly general terms thanks to the results of the previous sections, the attention will be focussed on the case where the slow-roll parameters are
approximately constant in time. 

According to Eq. (\ref{TT}) the  tensor modes only couple to the curvature and their evolution equations are always the same 
both in the adiabatic and in the quasiadiabatic case even if the basic fields driving the slow-roll dynamics change completely 
from one case to the other.  Equation (\ref{TT}) can be solved in the two relevant regimes namely for $k \tau \gg 1$ (when the 
relevant wavelengths are shorter than the Hubble radius) and for $k\tau \ll 1$ (when the wavelengths 
are larger than the Hubble radius). After counting properly the tensor polarizations, the power spectrum becomes, in terms 
of the rescaled variable $\mu_{k}(\tau)$
 \begin{equation} 
 P_{T}(k,\tau) = \frac{4 \ell_{P}^2\,\, k^3}{\pi^2\, a^2} |\mu_{k}(\tau)|^2,\qquad \mu_{k}^{\prime\prime} +[ k^2 - a^{\prime\prime}/a] \mu_{k}=0.
 \label{MF2}
 \end{equation}
The solution for $\mu_{k}$ in the two regimes $k\tau \gg 1$ and $k \tau \ll 1$ can be written as:
\begin{eqnarray}
\mu_{k}(\tau) &=& \frac{1}{\sqrt{2 k}} e^{ - i k\tau} , \qquad k \tau \gg 1,
\nonumber\\
\mu_{k}(\tau) &=& A_{k} a(\tau) + B_{k} a(\tau) \int^{\tau} \frac{d \tau^{\prime}}{a^2(\tau^{\prime})}, \qquad k\tau \ll 1.
\label{TSOL}
\end{eqnarray}
The values of  $A_{k}$ and $B_{k}$ appearing in Eq. (\ref{TSOL}) are determined by demanding the continuity in $\tau^{(T)}_{ex}$ 
of the solution and of its first derivative. We remind that, by definition, $k\tau^{(T)}_{ex} \simeq 1$. The full result, valid for $\tau \geq \tau_{ex}$ can also be written as
\begin{equation}
\mu_{k}(\tau) = \frac{e^{- i k\tau^{(T)}_{ex}}}{\sqrt{2 k}} \biggl[ \biggl(\frac{a}{a_{ex}}\biggr)  - a a_{ex} ({\mathcal H}_{ex} + i k) \int_{\tau_{ex}}^{\tau} \frac{ d \tau^{\prime}}{a^2(\tau^{\prime})}\biggr],
\label{TSOL2}
\end{equation}
where $a_{ex}= a(\tau_{ex}^{(T)})$ and analog notation is used for ${\mathcal H}_{ex}$. In the adiabatic case the large-scale curvature perturbations 
are caused by a single scalar field. As already mentioned in section \ref{sec3} the corresponding evolution equations can be recovered from our results by trivially setting $\overline{\xi} \to 0$, $\delta \xi \to 0$, $\delta p_{nad} \to 0$ and $c_{s} \to 1$ (see e.g. Eq. (\ref{evolR4})). Ultimately the well known evolution equation of the curvature perturbations will be given, in Fourier space, by
\begin{equation}
{\mathcal R}_{\varphi}^{\prime\prime} + 2 \frac{z_{\varphi}^{\prime}}{z_{\varphi}} {\mathcal R}_{\varphi}^{\prime} + k^2 {\mathcal R}_{\varphi} =0, \qquad z_{\varphi} = a \varphi^{\prime}/{\mathcal H}.
\end{equation}
In this case the power spectrum is given by 
\begin{equation}
P_{{\mathcal R}}(k,\tau) = \frac{k^3}{ 2 \pi^2 z_{\varphi}^2} |q^{(\varphi)}_{k}(\tau)|^2, \qquad q^{(\varphi)}= z_{\varphi} {\mathcal R},
\label{RSP}
\end{equation}
where $q^{(\varphi)}$ denotes the normal mode in the scalar field case while $q$ (without superscript) denotes the normal 
mode in the quasiadiabatic case (see, in particular, Eq. (\ref{MODEF1})).  The same procedure described in the case of the tensors and leading to Eq. (\ref{TSOL2}) can be applied in the case 
of $q^{(\varphi)}$. The full solution analog to Eq. (\ref{TSOL2}) but valid in the scalar case is given by:
\begin{equation}
q^{(\varphi)}_{k}(\tau) = \frac{e^{- i k\tau^{(S)}_{ex}}}{\sqrt{2 k}} \biggl[ \biggl(\frac{z_{\varphi}}{z_{ex}}\biggr)  - z_{\varphi} z_{ex} (z^{\prime}_{ex}/z_{ex} + i k) \int_{\tau_{ex}}^{\tau} \frac{ d \tau^{\prime}}{z_{\varphi}^2(\tau^{\prime})}\biggr].
\label{RSOL}
\end{equation}
Note that, in the case of Eq. (\ref{RSOL}) the solutions valid for $k\tau \ll 1$ and $k\tau \gg 1$ have been matched for $k\tau^{(S)}_{ex}\simeq 1$
where $\tau^{(S)}_{ex}$ coincides, in the genuine adiabatic case, with $\tau^{(T)}_{ex}$.
From these expressions it is therefore possible to compute the tensor to scalar ratio denoted by $r_{T}^{(\varphi)}$ which is 
given in the single scalar field case by:
\begin{equation}
r_{T}^{(\varphi)} = 8 \ell_{P}^2 \,\biggl|\frac{z_{ex}}{a_{ex}} \biggr|^2 = \frac{8}{ \overline{M}_{P}^2} \,\biggl( \frac{\dot{\varphi}^2_{ex}}{H_{ex}^2}\biggr) \,= \,
16 \,\epsilon \ll 1,
\label{TR1}
\end{equation}
where $z_{ex} = z_{\varphi}[\tau^{(S)}_{ex}]$. As expected 
the tensor spectral index is $n_{T} = - 2\epsilon$ to lowest order 
in the slow-roll approximation\footnote{This result is easily obtained by appreciating that in the evolution 
of the tensor mode functions $a^{\prime\prime}/a = a^2 H^2( 2 -\epsilon)$. Furthermore, since we are considering the case 
of constant slow-roll parameters (see Eq. (\ref{SR}) and discussion therein), $a H = -1/(1 - \epsilon)$. }.
When the fluctuations are induced by bulk viscosity the tensor contribution is 
exactly the same since what matters is the evolution of the extrinsic curvature which is only sensitive 
to the $\dot{H}/H^2$, at least during the slow-roll phase.  The evolution of the scalar modes has been already discussed in detail 
 in the last part of section \ref{sec3} (see, more specifically,  Eq. (\ref{MODEF1})  and discussion therein). 
Therefore from Eqs. (\ref{sol1})--(\ref{sol2}) and from Eqs. (\ref{sol1a})--(\ref{sol2a}) we can obtain the tensor to scalar 
ratio in the quasiadiabatic case:
\begin{equation}
r_{T} = \frac{8}{\overline{M}_{P}^2} \biggl| \frac{\overline{z}_{t}( \tau_{ex}^{(S)})}{a(\tau_{ex}^{(T)})} \biggr|^2 e^{ \frac{k^2 \tau_{ex}^{(S)}(1 + w)}{4 \epsilon}}.
\label{TSVISC}
\end{equation}
Naively from Eq. (\ref{TSVISC}) we could argue that $r_{T}$ gets much larger than $1$  in the limit $\epsilon \to 0$: this means 
that, in the quasi-adiabatic solution, that the scalar modes are 
suppressed in comparison with the tensor modes. Even this qualitative argument is grossly correct the situation is a bit more subtle since, as indicated, the conditions for the horizon crossing are different for the scalar and tensor modes. In particular, recalling  Eqs. (\ref{sol1})--(\ref{sol2}) and  Eqs. (\ref{sol1a})--(\ref{sol2a}), during the slow-roll phase we have that 
\begin{equation}
k \tau_{ex}^{(S)} \simeq 2 \sqrt{\epsilon}, \qquad k \tau_{ex}^{(T)} \simeq 1, \qquad a(\tau_{ex}^{(S)}) \simeq \frac{a(\tau_{ex}^{(T)})}{2 \sqrt{\epsilon}}.
\label{TV2}
\end{equation}
According to Eq. (\ref{TV2}) the evolution of the tensor modes implies that the horizon crossing occurs for $k \tau_{ex}^{(T)} =1$ while for scalars it occurs for $k \tau_{ex}^{(S)} = 2 \sqrt{\epsilon}$, i.e. $|\tau_{ex}^{(S)}| \ll |\tau_{ex}^{(T)}|$ since, during slow-roll,  $\epsilon \ll 1$.
With these specifications we have that Eq. (\ref{TSVISC}) can also be written as
\begin{equation}
r_{T} = \frac{16 \epsilon}{|c_{eff}|^2} e^{w + 1} \biggl[ \frac{a( \tau_{ex}^{(S)})}{a(\tau_{ex}^{(T)})} \biggr]^2 \simeq \frac{4}{|c_{eff}|^2}e^{w + 1}, \qquad \frac{r_{T}}{r_{T}^{(\varphi)}} \simeq \frac{e^{w +1}}{4 \, \epsilon \, |c_{eff}|^2}.
\label{TR2}
\end{equation}
If we compare Eqs. (\ref{TR1}) and (\ref{TR2}) we can observe that $r_{T}$ is at least ${\mathcal O}(100)$ times larger than 
$r_{T}^{(\varphi)}$. This estimate follows if we consider that, at most, $|c_{eff}|^2 < {\mathcal O}(1)$. Similarly we can take $w = {\mathcal O}(1)$ (but smaller than $1$). Note that this estimate does not rely on a particular inflationary solution 
driven by bulk viscosity but just on the assumption that slow-roll dynamics is compatible with a background viscosity.

In summary, if the quasi-de Sitter phase is caused by the evolution of the bulk viscosity the quasiadiabatic scalar mode is subleading in comparison with the tensor mode.  Conversely the bona fide adiabatic dominates against the tensor mode in the single field case. 
This result means that when the quasi-de Sitter phase is driven by bulk viscosity we are getting closer to the situation of the {\em exact} de Sitter space-time where the scalar fluctuations of the geometry should strictly vanish in comparison with the tensor mode. 

\renewcommand{\theequation}{6.\arabic{equation}}
\setcounter{equation}{0}
\section{Concluding remarks}
\label{sec6}
Depending on the specific dynamical situation the large-scale inhomogeneities of the viscous coefficients
 can be classified as entropic (i.e. nonadiabatic) or quasiadiabatic.  Whenever the bulk viscosity does not have a
homogeneous background the resulting fluctuations are automatically gauge-invariant
and their contribution to the evolution equations of the curvature perturbations reminds of the familiar source terms arising in connection with the
(four) conventional entropic modes customarily constrained by means of the temperature and the polarization anisotropies of the CMB. 
Along this first perspective, the viscous modes are only tolerable as a subleading component of a dominant adiabatic solution 
whose presence is instead mandatory in the light of current large-scale observations.  
A second complementary possibility stipulates that the viscous coefficients have a spatial variation 
but in the presence of a homogeneous background.  In such a situation the curvature perturbations inherit a source term depending on the 
fluctuations of the viscous coefficients, on the background viscosities and on the inhomogeneity of the expansion rate. 
The fluctuations of the bulk viscosity coefficient always act as a supplementary 
nonadiabatic pressure fluctuation but, this time, a quasi-de Sitter stage of expansion 
can be driven solely by the viscous coefficients. The present analysis  demonstrates that the evolution of curvature perturbations is in general nonadiabatic. Moreover, if and when the bulk viscosity coefficient leads to quasi-de Sitter solutions and to large-scale curvature perturbations, 
the shear viscosity coefficient at large scales only couples to the evolution of the traceless part of the extrinsic 
curvature of the spatial slices and does not contribute to the accelerated expansion. 

 The potentially dangerous nonadiabatic source terms summarized in the previous paragraph disappear whenever the viscosity coefficients depend solely on the energy density of the relativistic fluid. In this case the curvature perturbations are effectively 
quasiadiabatic since they coincide with the standard adiabatic solution but only in the large-scale limit.
In perturbation theory this conclusion follows from the analysis of the gauge-invariant fluctuations of the spatial curvature.
The same result can also be obtained from a fully nonlinear analysis where the evolution  
 of the curvature perturbations is studied within the gradient expansion appropriately extended to handle 
 the viscous situation.  Unfortunately  the curvature power spectrum of the quasiadiabatic solution is parametrically smaller than 
 the tensor power spectrum. Hence the corresponding tensor to scalar ratio turns out to be larger than in the 
 standard adiabatic case where the scalar power spectrum dominates, over large scales, against its tensor counterpart.  
Taken at face value the obtained results show that the viscous coefficients alone cannot drive a phase of accelerated expansion and, at the same time, reproduce the standard adiabatic scalar mode. We can get very close to an acceptable phenomenological situation if the bulk viscosity 
coefficient only depends on the energy density of the plasma. Even in this case, however, there are serious drawbacks since 
the dominance of the tensors against the scalars is at odds with a pretty robust observational evidence. In this respect 
the obtained results suggest a novel strategy for a concrete phenomenological scrutiny of the large scale inhomogeneities induced by the viscous coefficients.

In a more optimistic perspective the large-scale fluctuations of the viscous coefficients remain a viable possibility only when the dominant adiabatic solution comes from a different physical origin. In this case the fluctuations of the bulk viscosity play the same r\^ole of a supplementary nonadiabatic solution in the space of the initial conditions of the Einstein-Boltzmann hierarchy. Such a component can be constrained prior to photon 
decoupling and across the matter-radiation transition with the same techniques customarily employed 
to bound the presence of the standard four nonadiabatic solutions.  
\section*{Acknowledgments}
It is a pleasure to thank T. Basaglia and J. Jerdelet of the CERN scientific information service for their kind assistance.

\end{document}